# Explaining Autonomous Decisions in Swarms of Human-on-the-Loop Small Unmanned Aerial Systems

**Ankit Agrawal, Jane Cleland-Huang**

Computer Science and Engineering
University of Notre Dame, Indiana, USA.
aagrawa2@nd.edu, janehuang@nd.edu

## Abstract

Rapid advancements in Artificial Intelligence have shifted the focus from traditional human-directed robots to fully autonomous ones that do not require explicit human control. These are commonly referred to as Human-on-the-Loop (HotL) systems. Transparency of HotL systems necessitates clear explanations of autonomous behavior so that humans are aware of what is happening in the environment and can understand why robots behave in a certain way. However, in complex multi-robot environments, especially those in which the robots are autonomous, mobile, and require intermittent interventions, humans may struggle to maintain situational awareness. Presenting humans with rich explanations of autonomous behavior tends to overload them with too much information and negatively affect their understanding of the situation. Therefore, explaining the autonomous behavior or autonomy of multiple robots creates a design tension that demands careful investigation. This paper examines the User Interface (UI) design trade-offs associated with providing timely and detailed explanations of autonomous behavior for swarms of small Unmanned Aerial Systems (sUAS) or drones. We analyze the impact of UI design choices on human awareness of the situation. We conducted multiple user studies with both inexperienced and expert sUAS operators to present our design solution and provide initial guidelines for designing the HotL multi-sUAS interface.

## Introduction

In traditional human-in-the-loop systems, a human interacts closely with the system to make plans, approve actions, and serve as the primary decision-maker, while the system assumes responsibility for enacting those human-initiated plans (Nunes, Silva, and Boavida 2018). In this scenario, the human knows what tasks the system is performing and why it is performing them. In the rapidly emergent arena of human-on-the-loop (HotL) systems (Fischer et al. 2017), the system is imbued with the ability to make and enact its own decisions. However, in HotL systems, humans must still perform a supervisory role. They therefore need to maintain Situational Awareness (SA), defined as the ability to perceive the environment (Level-1), understand the reasoning behind the current state of the environment (Level-

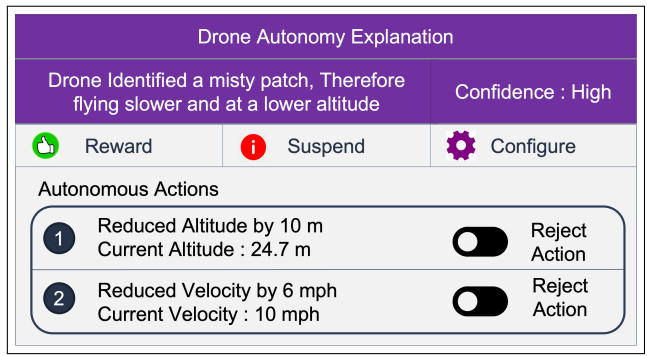

Figure 1: This detailed explanation of a change in the sUAS flight pattern indicates that the sUAS encounters misty weather conditions and adapts its flight pattern. The explanation includes the sUAS response (flying lower and slower), planned operational changes (velocity and altitude), and cause of this change (mist identification).

2), and finally, to project how the situation could evolve in the future (Level-3) (Endsley 1995b).

Human intervention in HotL systems takes many forms and requires the human operator to maintain SA so that they are able to intervene when needed - typically through creating a new plan, overriding a decision, reverting permissions, or assuming temporary control (Agrawal, Steghöfer, and Cleland-Huang 2020). Therefore, the human interface of the HotL systems must be carefully designed to give human operators time to intervene as well as mechanisms for supporting the intervention. Even with such affordances, HotL systems tend to suffer from three well-known human-machine interaction problems due to their highly autonomous nature. These include (a) automation bias, in which a human places excessive trust in the machine's decision-making, (b) loss of SA, and (c) shifting of moral responsibility to the machine (Challen et al. 2019). Communicating the reasoning behind the autonomous decisions to humans through a human-machine interface can alleviate these problems since humans can assess the robot's reasoning and evaluate where it has taken appropriate actions in a particular scenario and context (Koo et al. 2015). Therefore, HotL systems rely heavily on generating explanations of autonomous behavior so that humans can understand what is happening in the environment and why it is happening (Wortham, Theodorou, and Bryson 2017). Such explanations allow human operators to understand autonomous actions and develop a multi-faceted conceptual model of the system's autonomy. This conceptual model includes their understanding of the system's capabilities to perform correctly under diverse conditions, its reliability for correctly completing a task (Dixon and Wickens 2006), and the degree of human trust that can be placed in the system (Rogers et al. 2019; Schrills and Franke 2020). All of these factors govern human intervention behavior in a HotL system, and suggest the importance of providing rich and timely explanations of autonomous behavior.

Humans supervising multiple small Unmanned Aerial Systems (sUAS) in emergency response missions achieve SA through directly observing the actions, location, and health of each sUAS in flight and also through information provided by the User Interface (UI). Humans also need to understand rationales behind the sUAS' autonomous decision-making (Li et al. 2020) so that they can make judgments about the correctness of the sUAS' behavior. However, adding too much explanatory material with too many details about the autonomous behavior can lead to the well-known SA design demon of 'information overload' (Klingberg 2009). We therefore explore design solutions for explaining autonomous decisions of multiple sUAS to a human operator in order to enhance their SA while supervising a HoTL system and intervene when needed.

In this paper, we describe a study that we conducted in the domain of sUAS, as they represent a rapidly emergent area of HotL systems with diverse application domains such as medical delivery (Mesar, Lessig, and King 2019; Claesson et al. 2017), multi-sUAS area search (Scherer et al. 2015), ice rescue (Iob et al. 2020), and fire surveillance (Wakeham, Griffith, and Campus 2015; Khan and Neustaedter 2019). sUAS applications often involve multiple human operators and multiple sUAS – collaborating together to achieve a specific task (Agrawal et al. 2020; Cleland-Huang and Agrawal 2020). Our study explored human interface design tensions in explaining the autonomy of aerial search using multiple autonomous sUAS. We focused particularly on two of the known challenges of autonomy in human-machine interfaces, namely (1) loss of SA (Endsley 2017) – with an emphasis upon understanding sUAS autonomy, and (2) automation bias, defined as the propensity for human operators to accept suggestions made by automated decision-makers without question, while ignoring available contradictory evidence. Our study was designed to address the following research questions:

**RQ1** How do simultaneous explanations of multiple sUAS behaviors impact remote operators' SA?

**RQ2** How does the explanation of autonomous behavior influence automation bias in a multi-sUAS environment?

**RQ3** What UI design trade-offs arise from explaining autonomy in a multi-sUAS environment, and how can we resolve these trade-offs while providing sufficient SA to remote operators?

Based on the analysis of the data collected during our study, we provide initial guidelines on trade-offs that should be considered in the design of a multi-sUAS UI. Our initial recommendations include (1) adapting the level of detail in the explanations (2) excluding operational details from the explanations, (3) favoring the use of icons and animations over textual descriptions as observed by Lester et al., (Lester 2013), (4) providing affordances that allow users to validate the information provided in the explanation, and (5) prioritizing explanations of autonomous behavior that could jeopardize the mission's objective. We describe these recommendations in detail later in the paper.

The remainder of the paper is laid out as follows. First, we discuss strategies for explaining sUAS autonomy, designing AI-driven user experiences, and SA design demons. Second, we describe our DroneResponse platform on which we conduct our analysis. Third, we describe our initial design of explanations, summarize the feedback obtained from preliminary participatory design exercises, and describe the subsequent improvements made to the UI. Fourth, we describe the user study design and experiments conducted to evaluate the effectiveness of different techniques for explaining sUAS autonomy. Fifth, we report results from our user studies and propose subsequent design improvement, and then finally, we discuss the study's limitations and conclusions.

# Background

The explainability of intelligent systems and the impact of explanations upon the user's mental model have been widely studied in various contexts including machine learning classifiers & AI models (Chakraborti et al. 2017), recommendation systems (Dominguez et al. 2019), and planning & decision systems (Kasenberg, Thielstrom, and Scheutz 2020; Zakershahrak et al. 2020). In the context of autonomous agents, Titarev et.al (Tintarev and Kutlak 2014) developed Scrutable Autonomous Systems (SAsSy) to demonstrate a human-understandable dialog explaining an autonomous agent's behavior through argumentation and natural language. Garcia et al.,(Garcia et al. 2018) designed a natural language UI to provide explanations of autonomous behavior to remote operators. Both systems provided explanations in the form of dialogue upon human request. Our work closely aligns with these approaches through explaining autonomous behavior. However, due to the high autonomy levels and the time-sensitive nature of sUAS flight missions, our UI design pushes explanations of autonomous behavior to the operators rather than requiring humans to request explanations upon demand. Previous studies have shown that the use of graphics such as icons and images for visual communication (Lester 2013) is fast, engaging, and attention-grabbing. Therefore, we combined both graphical and natural language modes of explanation to design our UI for explaining sUAS autonomy in a time-constrained environment.

### AI-Driven User Experiences

Explanations in a multi-sUAS mission domain must be formatted in a way that supports *immediate* and *clear* comprehension, as misinterpretations may have safety and legal consequences. Since the autonomy of sUAS is dependent on the ability of the AI models to recognize their surroundings, it is important to examine how wrong decisions and their explanations are perceived by humans in a complex and time-constrained environment (Kliman-Silver et al. 2020). Therefore, in our study, we deliberately included scenarios (see Table 1) in which one or more sUAS presented information as if they had incorrectly perceived the environment (Scenarios V1, V3, and V4), and also cases in which they failed to execute correct actions during the mission (Scenario V5). This allowed us to evaluate the SA of users under such conditions.

Gregor et al. (Gregor and Benbasat 1999) investigated the nature and usage of explanations across diverse autonomous systems. They showed that explanations should address *"Why"* and *"Why not"* a certain decision is chosen over others. For self-driving cars, Koo et al., (Koo et al. 2015) also found that explaining "Why" was essential for maintaining good driving performance. Therefore, in this paper, we focus on explaining *why* an sUAS is acting in a certain way through our multi-sUAS interface. For example, an sUAS starts flying lower and slower *because* misty weather is detected, or an sUAS switches from searching to tracking mode *because* it recognized a person on the ground.

### Situational Awareness (SA)

A design tension exists between providing a detailed explanation of sUAS autonomous decisions with the overall need for the human supervisor to maintain SA of the mission. Endsley previously identified eight common design problems impacting SA (Endsley 1995b) and referred to them as Design Demons. Prior work has suggested that four design demons are particularly relevant to multi-sUAS scenarios (Agrawal et al. 2020). These include (1) Information Overload (IO), which occurs when the human operator faces difficulties in processing all the presented information, (2) Errant Mental Model (EMM), in which the human forms an incorrect mental model of the situation, (3) Misplaced Salience (MS), characterized by inappropriate use of colors, alarms, and warnings prohibits that make it difficult for users to make connections between different pieces of information, and (4) Attentional Tunneling (AT), in which the user fixates on a single piece of information at the expense of missing critical information.

## DroneResponse: Multi-sUAS Search

We utilized the UI and executable environment of the DroneResponse system (Agrawal et al. 2020; Cleland-Huang et al. 2020) to experimentally evaluate the efficacy and trade-offs of different design decisions for explaining autonomy. DroneResponse uses physical sensors such as a camera, GPS, and LiDAR to sense the environment, and leverages onboard computer vision (CV) models (e.g., YOLOv3 (Redmon et al. 2016)), to analyze data and to identify specific objects. The autonomous behavior of each individual sUAS is determined by an internal state-transition model, which tracks its current state (e.g., searching, tracking), and specifies the permitted state transition conditions. Multiple sUAS communicate over secured Wi-Fi or LTE channels with the DroneResponse internal air-traffic control (ATC) service whenever synchronization is required. A dedicated video streaming server interfaces with each sUAS over WiFi/LTE to stream close to real-time video from sUAS to the UI.

### sUAS Autonomy

For the study, we identified four common types of incidents that require sUAS to adjust their flying patterns. Each incident is described below:

- **Weather Recognition**: The sUAS detects local and transient weather conditions including mist, snow, and rain using a weather recognition CV model (Abraham et al. 2021). This ability is important for vision-based search, as inclement weather impacts visibility and requires the sUAS to adapt its flying behavior. For example, if an sUAS detects rain while searching for a victim, it should autonomously reduce its altitude and velocity (i.e., fly lower and slower) to preserve image detection accuracy.

- **Autonomous Tracking**: During a search, the sUAS uses onboard CV to detect a person in the video frame. Once a person is detected with sufficient confidence, the sUAS switches into tracking mode and navigates in 3D space according to the movements of the person.

- **Return to Launch**: sUAS autopilot software, provides a fail-safe mechanism to return the sUAS to its launch coordinates (RTL) when the battery level drops below a predefined threshold level or when signal is lost. The autopilot provides the ability to override this fail-safe, so that the sUAS identifies and lands at the closest predefined landing pad.

- **Dynamic Path Planning**: Each sUAS needs to perform path-finding to avoid no-fly zones and obstacles. While DroneResponse is not yet fully implemented with dynamic navigation, we have created a mock-up in the UI, so that UAVs avoid no-fly zones, such as hospitals, stadiums, and universities. The sUAS are programmed to select the shortest direct route, or to request human assistance for planning if they are unable to find a viable path to their destination.

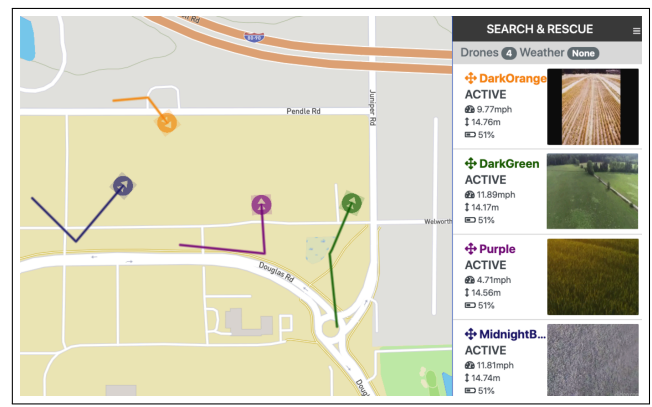

Figure 2: DroneResponse UI showing flight paths of four sUAS. sUAS information and their icons on the map are color-coded.

## Prototype Design

In previous work, we conducted a participatory design session with stakeholders to collect end-user requirements and proposed interfaces for multi-sUAS systems (Agrawal et al. 2020). We developed a high-fidelity interactive UI of our initial design as shown in Figure 2, and then augmented the design to include explanations of autonomous sUAS behavior, which is the focus of this study. We considered the autonomy scenarios described in the section on "sUAS Autonomy" to generate design ideas for explaining the sUAS autonomy.

In each scenario, we identified the necessary pieces of information for understanding the autonomous behavior of sUAS. These included the *event* that occurred (e.g., mist was detected) causing the sUAS to respond autonomously, the *action* taken by the sUAS in response to that event (e.g., flying lower and slower), and finally the *reasoning* behind the decision (e.g., to increase visibility) (Mualla et al. 2020). In addition, we considered supporting the explanation of sUAS autonomy by describing *operational level changes*, such as a change in the flight mode or flight plans, and also by presenting the sUAS confidence in their AI-driven autonomous decisions (Zhang, Liao, and Bellamy 2020). After exploring multiple options for embedding the explanations concisely in the UI, we decided to provide event recognition information immediately next to the sUAS's current location icon in the UI, as shown in Figure 3. We reasoned that this design would allow remote pilots to maintain awareness of events in the environment, and understand which sUAS are responding to those events when multiple events occur simultaneously.

Finally, we also developed an explanation box to provide details of sUAS autonomous behavior (see Figure 1) and color-coded it with the sUAS icon on the map for easy comprehension. The explanation box included the autonomous actions, reasoning, and operational changes in flying behavior. Further, we included interactive options for remote pilots to interact with sUAS autonomy, allowing them to configure, suspend, and acknowledge the sUAS' autonomous behavior. When an event occurs in an environment, UAV explanations appear on the UI without the user having to request them explicitly. This design supports the human-on-the-loop interaction style of the system.

## Design Exercise

We conducted an initial design exercise with senior undergraduate students who were taking a class on software development for sUAS because they were familiar with sUAS operations. The purpose of this exercise was to collect preliminary feedback, refine our prototype, and identify challenges that remote pilots might encounter in comprehending a situation. 26 of 30 students chose to participate in the design exercise, which involved reviewing screen recordings of simulated multi-sUAS missions where one or more sUAS adapts and explains their autonomous behavior. The students individually described what they understood to be happening in each scenario and provided suggestions for improvements in the design.

Within each of these scenarios, sUAS video streams were provided by prerecorded video clips (collected partially from YouTube and partially using our own sUAS). Adobe Premiere was used to add weather effects to the clips, and YOLO the (Redmon et al. 2016) vision model was used to detect people and to draw bounding boxes around them. Finally, we simulated the sUAS in Gazebo (Koenig and Howard 2004) and executed complete missions in DroneResponse, recorded them using screen recording tools, and presented them to students during the design exercise. It is worth noting that the DroneResponse UI is identical regardless of whether simulated or physical sUAS are flown. We included at-least one video for each of the four autonomous events described previously. Each event was observable in the UI via the video stream of at least one sUAS and information available on the map. We also included scenarios where an sUAS reports an incorrect event. For example, in one scenario, the sUAS mistakenly interpreted a ball as a person and then started tracking it.

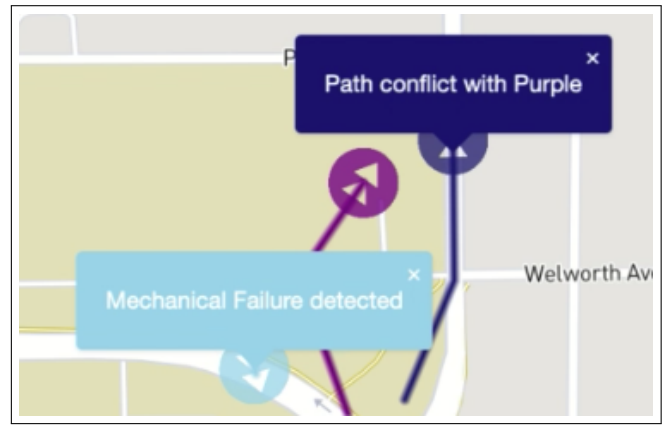

Figure 3: Multiple sUAS are reporting the event on the UI as it occurs

## Design Improvements

We reviewed the feedback to identify potential opportunities for design improvement. As a result, we modified the way in which autonomous sUAS behavior was explained on our UI. We categorized these design changes into four design concepts: *Information Clustering*, *Uncertainty Prioritization*, *Event-Action Pairs*, and *Failure Avoidance* to improve the SA of sUAS operators. We briefly describe some preliminary feedback.

**Information Clustering:** One person reported that the pieces of information needed to completely understand the explanation were scattered across the screen and suggested bringing the drone's camera view to the user's attention by enlarging or highlighting it when an event occurred. This comment matched one made in a previous study (Agrawal et al. 2020), where firefighters stated that when they detected a victim, they wanted to focus all of their attention on the victim. Therefore, we redesigned our detailed explanation by moving relevant information inside a single explanation box. In addition to describing the operational changes in the sUAS flying pattern, we added video frames that were rel-evant to the AI decision-making into the explanation box. This change allowed the users to validate the sUAS actions without switching attention between different areas of the screen. Figure 4a shows the screenshot of the refined prototype, based on the hypothesis that it would require less time for a user to validate events, and that SA could be maintained more effectively, if relevant information were made available in a single place.

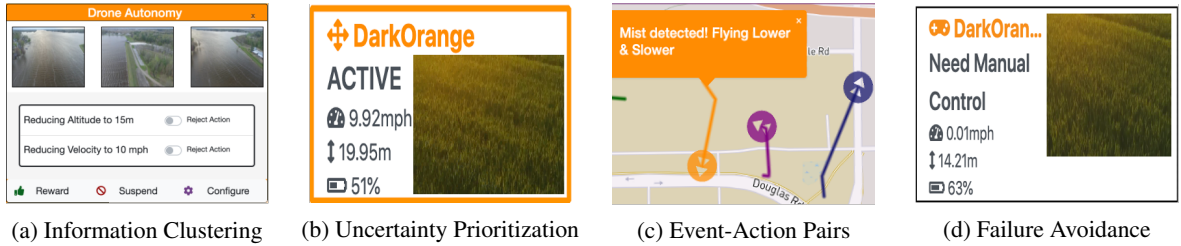

(a) Information Clustering  (b) Uncertainty Prioritization  (c) Event-Action Pairs  (d) Failure Avoidance

Figure 4: Design Concepts applied to manage the information overload while providing multi-sUAS autonomy explanations **(a)** Group relevant information including a description of the event and video frames providing evidence supporting the sUAS autonomous decision. **(b)** Explicitly highlighting the border of the information box of the sUAS that is exhibiting autonomous actions to draw user's attention. **(c)** Presenting "Event" and "Action" together as a pair. **(d)** Blinking icon and text is used when human control is required in an emergency situation. In this case, the text "Needs Manual control" and the handheld controller icon starts blinking to draw immediate attention.

**Uncertainty Prioritization:** Information overload was clearly evidenced, as 83% of the students were unable to understand or validate all the events that occurred simultaneously. To address this phenomenon, we modified the design to only display a detailed explanation (i.e., the larger explanatory box as depicted in Figure 4a) when sUAS confidence in the autonomous decision was below a predefined threshold level. When an explicit explanation was omitted, we colored the border of the sUAS information box with its uniquely assigned color, to serve as an indicator that the sUAS is currently executing an autonomous action. In this way, we used visual communication to draw the participant's attention to the sUAS status and to its video stream in the right-side panel of the screen. This design concept is illustrated in Figure 4b. Table 1 indicates the scenarios for which explicit explanations were omitted from events due to high sUAS confidence in their autonomous actions. These scenarios were specifically crafted to evaluate whether we could use this technique to reduce information overload while maintaining sufficient SA of the remote operators.

**Event-Action Pairs:** Multiple participants mentioned issues in quickly parsing all the explanatory information and suggested that more concise event pop-ups might be helpful. In order to provide information concisely and to address the Errant Mental Model design demon, we moved the textual explanation of the event and sUAS(s) adaptations in response to that event from the explanation box to the pop-up which is adjacent to each sUAS' location as shown in Figure 4c. Placing a brief description of the sUAS' autonomous behavior next to the sUAS reduces the likelihood that users will confuse one sUAS' autonomous actions with another's.

**Failure Avoidance:** Efficient Human-Drone partnerships require humans to intervene in the sUAS autonomy when the sUAS makes incorrect decisions or when the sUAS proactively seeks human assistance. In one of our scenarios, the sUAS explicitly requested help from the human. The following message was displayed immediately below the autonomy explanation: *"RTL command failed, need human control to prevent crash"*. To raise human awareness that manual take-over was required, we replaced the sUAS *Active* status with *Needs Human Control* as displayed in the right side panel of the screen. Furthermore, we included a blinking handheld controller icon to reinforce the message. Figure 4d shows the prototype after making these changes.

## Experimental Design

After prototype refinement, we conducted user studies that included both inexperienced sUAS operators and experts to capture their SA under different situations. The goal of the user study was to better understand how our design decisions influenced the SA of remote pilots and other observers under different real-time situations. Overall, we sought to understand the strengths and limitations of each design concept, and which ones were better suited for different mission contexts.

Each recorded mission scenario (referred to as a 'mission video' from now on) used in our study, was around 90 seconds, and included one or more sUAS modifying their flying behavior in response to various environmental events. Scenario characteristics are provided in Table 1. In mission V1, an sUAS incorrectly believes an object on the ground to be a person and starts tracking it, while in V2, an sUAS correctly recognizes the misty weather conditions and adapts its flying patterns. The mission videos V3, V4, and V5 include simultaneous explanations from two sUAS. In V3 and V4, one sUAS out of two fails to recognize an event correctly, while in V5, one sUAS executes the actions properly but requests human assistance. Irrespective of the number of sUAS acting autonomously, the UI always provided an explanation reflecting the sUAS' belief as illustrated in Figure 4.

### User Engagement

We recruited MTurk workers, without sUAS expertise (non experts), to watch the videos and then answer a set of questions about their observations and perspectives. In addition, we engaged seven people with experience flying autonomous aerial vehicles including certified FAA pilots, sUAS hobbyists, and researchers working on sUAS development, in order to potentially obtain richer feedback and to determine whether the perception of sUAS autonomy differs between non experts and domain experts. In both studies, we elicited feedback from the participants using mission videos based on the modified prototype produced as a result of the preliminary participatory design process.

### Study Format

Each participant watched at least one mission videos, and after each video was asked to respond to a questionnaire to document their awareness of the situation. Following the Situational Awareness Global Awareness Technique (SAGAT) (Endsley 2000, 1995a), when participants were observing the mission, we froze the video after an sUAS explained its behavior. Participants were encouraged to describe their overall impression of the UI, autonomy, and any challenges in understanding what was going on in each mission. We then performed a quantitative and qualitative analysis of the collected responses in order to answer our primary research questions with a focus on the following three factors.

- **sUAS Autonomy Awareness**: Each participant was asked to describe their understanding of the mission in writing. We analyzed each textual response to identify 1) their awareness of the sUAS' autonomous actions, 2) their perception of the environment, 3) how they dealt with attentional challenges whilst monitoring multi-sUAS' autonomous activities in the time-constraint environment, and finally 4) how each of the individual design decisions helped them to perceive the mission.

- **Human Automation Bias**: Given the well-known human-machine problem of *automation bias*, we sought to understand how an operator perceives the situation when a sUAS provides incorrect information. In scenarios in which the sUAS provided incorrect information, or adjusted their behavior based on events that they had misinterpreted (e.g., claimed to have detected mist, even though no mist existed), we evaluated whether users realized that the sUAS had misinterpreted its environment or not. Based on this analysis, we identified the user's propensity for accepting sUAS explanations, which in turn allowed us to evaluate the impact of automation bias and answer our third RQ.

- **Human-Agent Partnerships** : When humans work in multi-sUAS environments, they need to monitor the behavior of all participating sUAS, and develop an accurate mental model of the mission status. Humans can better collaborate with sUAS and assist them when necessary if they are fully aware of the environmental conditions in which each sUAS is operating. In our mission videos, the sUAS' environment is portrayed through its status (e.g, battery, location) and its video stream. Therefore, we observed the way users inspected, and responded to, the sUAS aerial video streams and the information displayed on the map, in order to analyze their ability to supervise the sUAS.

## Results and Analysis

### Procedure

Participants were first asked for their informed consent. Second, they performed a screening test to evaluate their ability to observe details and to correctly answer two questions about an image. Participants who failed the screening were eliminated from the study in order to improve data quality. Third, we provided participants with an overview of the project and familiarized them with the UI of the DroneResponse application by presenting and explaining several screenshots. We also explained the format of the survey, the estimated completion time, average length of each mission video, and the number of videos they would watch. In each case, we asked participants to assume the role of an emergency responder (e.g., a fire-fighter) responsible for monitoring the aerial search. Participants recruited from MTurk were randomly assigned one video clip each (V1, V2, V3, V4, V5) to keep the duration of the study shorter. However, domain experts watched all five videos, presented in random order. Finally, a post-task survey was administered following the SAGAT principles to assess participants' SA and to investigate if there were any indications of human-agent collaboration or automation bias.

### Participants

We recruited crowd workers without sUAS operational experience through MTurk. Each worker had prior task approval ratings of at least 99 percent and a minimum of 50 tasks completed. We collected data from 109 crowd workers who passed our pre-qualification test. To ensure data quality, we discarded two responses because the respondents did not watch the complete video during the study, 31 for providing vague or irrelevant responses, and one for being incomplete. Finally, we analyzed data from 78 participants (25 Females, 53 Males). Participants completed the entire study within an average of 13.29 minutes and were paid $2 for completing the task (equivalent to an hourly wage of $9.02).

In addition to the MTurk workers, we also recruited domain experts by sending emails to people in our personal networks whom we knew to be knowledgeable about sUAS. These experts included FAA remote pilots, sUAS developers, and people working on sUAS projects, as shown in Table 2. We analyzed a total of 113 scenario responses, including 78 from MTurk workers and 35 (5X7) from experts.

### Qualitative Analysis

We focused our qualitative analysis on understanding the remote operator's perception of the multi-sUAS mission, including challenges they faced whilst observing the mission, and their awareness of the sUAS autonomy and the environment. Two authors of this paper systematically applied inductive coding (Thomas 2006) to identify primary challenges that participants' face while observing and understanding the information provided in the multi-sUAS mission. We identified comments in the textual response indicating an issue, and repetitious comments about the same issue from a single participant were only counted once. In total, we observed 13 unique challenges, which we grouped into six primary themes as described in Table 3.

|  |  | **Misty Weather** | | | **Person Detected** | | | **Mechanical Failure** | | | **Path Re-planning** | | |
|---|---|---|---|---|---|---|---|---|---|---|---|---|---|
| ID | Count | Real Event | sUAS Action | Minimal Explain | Real Event | sUAS Action | Minimal Explain | Real Event | sUAS Action | Minimal Explain | Real Event | sUAS Action | Minimal Explain |
| V1 | Single |  |  |  | ○ | A2 | ✓ |  |  |  |  |  |  |
| V2 | Single | ● | A1 | ✓ |  |  |  |  |  |  |  |  |  |
| V3 | Multiple | ○ | A1 | ✓ | ● | A2 | ✓ |  |  |  |  |  |  |
| V4 | Multiple | ○ | A1 |  | ● | A2 |  |  |  |  |  |  |  |
| V5 | Multiple |  |  |  |  |  |  | ● | A6 |  | ● | A4 |  |

Table 1: A: Action; A1: Fly Lower and Slower, A2: Track, A3: Return to Home, A4: Path Re-plan, A5: Do Nothing, A6: Request Control. ● Event occurred.   ○ sUAS reported event, but it didn't actually occur.   ✓Detailed explanation was not provided due to high confidence in sUAS decision. The videos are available on YouTube [1].

| Ps | sUAS Exp. | Area of Expertise |
|---|---|---|
| P1 | 5 years | Professional sUAS developer |
| P2 | 4 years | Experienced sUAS pilot |
| P3 | 1 year | Researcher on sUAS vision. |
| P4 | 1 year | Researching CV with sUAS. |
| P5 | 2 year | Researching CV with sUAS. |
| P6 | 2 years | Professional sUAS developer |
| P7 | 3 years | Pilot and owner of sUAS startup |

Table 2: Details of Domain Experts participate in our study

The most frequent comment focused on *"things happening too fast"* which made it difficult for users to fully understand the state of the mission. One MTurk participant mentioned that *"I just saw that the drones were all doing separate things and too quickly for me to understand"*, while another reported *"There was so much going on at once that it was very stressful and difficult to watch all four drones at once"*. Domain experts also experienced time pressure as indicated by P4's comment that *"I only had time to read the event details for the purple drone at the bottom of the screen"*, and P2 who mentioned that *"Too many events happening at the same time, shortly after each other - hard to read and comprehend everything that happened."* Participants clearly suffered from Information Overload that hampered their decision-making. They offered comments such as *"there seemed to be too much information going on in the screen at the same time for me to evaluate and make any decisions"*. We observed similar comments from domain experts, as P5 specifically mentioned *"Since there are a lot of things happening on screen it can be hard at first to know what to focus on"*, and P7 also mentioned that *"when the alerts popped up, the information was slightly overwhelming to process quickly"*.

Another critical challenge was to associate the sUAS activity on the map with its corresponding information on the side panel based on colors and confidently confirm or reject the sUAS perception of the environment in a limited time. For instance, one MTurk participant reported *"I feel like I missed things as I moved from left [Map View] to right [Video Stream View]"*. Additionally, we saw that participants prioritized their visual attention on the events that mattered most to them. For instance, one MTurk participant focused on the map to make sure two nearby sUAS did not collide while failing to validate an event reported by one of the sUAS. They stated that *" One drone gave a notice that there was mist in the area... I checked the drones' positions to make sure that when they were in the flight path of another drone that their altitudes were different"*. Domain expert P1 intrinsically focused their attention on maps to better understand the situation and mentioned *"my eyes were drawn to the map to see where the drones were moving and I didn't see ..."* Furthermore, to alleviate the issue of Information Overload, domain expert P4 suggested focusing the explanation on critical events and mentioned *"The weather condition detection doesn't seem to play a very important role here (in the display) and can be confusing as an alert"*. These comments are indicative of a complex and rapidly changing environment.

### sUAS Autonomy Awareness

When only one event occurred in the environment (V2), and the sUAS correctly responded to it, we found that everyone was aware of the sUAS' autonomous action and the rationale for taking that action. Furthermore, in scenario V1, 87.5% of crowd workers demonstrated awareness of the sUAS autonomous action when it responded incorrectly to an event that never actually occurred in the environment. However, both domain experts and crowd workers experienced greater difficulty in perceiving the environment and the sUAS' autonomous actions when two events occurred simultaneously.

First, by comparing data collected from MTurk for scenarios V3 and V4, which both included simultaneous events, we observed that 64% of participants were aware of both events and understood the corresponding sUAS actions when a minimal explanation was provided in V3, but only 48% achieved this when the detailed explanation was provided in V4. In other words, providing a minimal explanation resulted in approximately 16% improvement in comprehen-sion.

---
[1] https://tinyurl.com/UAV-Autonomy-Study

| Themes | Total | Mission Scenarios | | | | |
|---|---|---|---|---|---|---|
| | | V1 | V2 | V3 | V4 | V5 |
| 1. Things happening too fast | 22 | 1 | 2 | 5 | 7 | 7 |
| 2. Difficulty in recognizing objects in video stream | 15 | 10 | 1 | 1 | 3 | 0 |
| 3. Need to prioritize visual attention | 14 | 1 | 5 | 0 | 7 | 1 |
| 4. Too much simultaneous information | 8 | 0 | 0 | 2 | 2 | 4 |
| 5. Difficulty switching between map view & information panel | 4 | 0 | 0 | 2 | 1 | 1 |
| 6. Feeling stressed | 2 | 0 | 0 | 2 | 0 | 0 |

Table 3: Autonomy awareness themes that emerged from the qualitative analysis of MTurk participants' feedback

Second, we observed that participants noticed at least one of the events, with 88% of participants in V3 demonstrating complete awareness of the "Person Detection" event, and 87% of participants showing awareness of the "Mist Detection" event in V4. while domain experts also prioritized their attention on a single event when two events occured in V3, P1 demonstrated complete awareness of both the autonomous action and its detailed explanation. Domain experts were more informed in scenario V3 where six out of seven of them showed awareness of both autonomous actions that had occurred – even though they did not know the operational details. In this regard, minimal explanations helped both experts and inexperienced operators to gain a better understanding of sUAS autonomy. When the events and actions were displayed as a pair adjacent to the sUAS's location on the map, the lower cognitive load helped people to comprehend both events in the environment. Also, the absence of additional operational details reduced the number of potential distractions and allowed the participants to focus on the actual mission. Finally, we observed that the detailed information in V4 was not as helpful since only 25% recognized the shift from "Search" to "Track" after the sUAS detected the person. Therefore, we concluded that removing operational details from the explanations enabled humans to better focus on multiple events and to make sense of the situation at hand. On the other hand, when sUAS performed correct autonomous actions without making any mistakes (V5), 70% of the participants demonstrate awareness of both autonomous actions suggesting that the automation bias, discussed next, was a contributing factor.

Hence, in response to **RQ1**, we found that providing detailed explanations of the autonomous behavior of multiple sUAS actually reduced the ability of users to maintain adequate SA in our studied time-critical environment. Furthermore, our results suggest that explanations in the UI should provide minimal information needed by the remote pilot to make an immediate decision rather than providing detailed reasoning for the autonomous decision of sUAS.

### Human Automation Bias Analysis

In our post-task questionnaire [2], we asked MTurk participants a series of multiple-choice questions regarding their understanding of the situation. To gauge whether participants were aware of the autonomous actions of sUAS, we counted the correct number of responses provided with respect to their observations about environmental conditions that triggered autonomous decisions, changes in sUAS velocity and altitude, and fairness of sUAS autonomous actions to assess diverse states of the mission. On average, MTurk participants only perceived 23% of the actual mission status correctly when sUAS made an incorrect decision (V1) as compared to 62.5% when the sUAS acted correctly (V2). This suggests that participants demonstrate the tendency to agree with the explanations and believed that the sUAS acted correctly. Similarly, when multiple events occurred, the participants were able to perceive 49% of the mission status when both sUAS notified the occurrence of the event correctly (V5) as compared to 39.5% (V3) and 32% (V4) when only one sUAS reported a correct event. We found that domain experts also suffered from Automation Bias, evidenced by the fact that in scenario V1, five experts incorrectly accepted the sUAS autonomous action and believed that there was a person present in the scene.

Hence, In response to **RQ2**, we found that explanations exacerbated the issue of automation bias. This observation confirms prior research results that show that explanations tend to increase the human's trust in the AI system, whether it is correct or not,(Lee and See 2004), making them more likely to follow the AI recommendation. While these findings are consistent with the psychological literature (Koehler 1991), we believe that other factors, such as time constraints in multi-sUAS environments exacerbate the situation, further hampering human ability to perceive and evaluate information correctly.

### Human-Agent Partnership Analysis

We analyzed user behavior towards inspecting video streams or available environmental information on the map to confirm or refute the validity of explanations presented to them. We found that most MTurk participants did examine the video stream or mentioned wanting to look at the video stream to validate events when only a single event took place (81% and 83% for V1 and V2 respectively). In contrast, when multiple events occurred in V3 and V4, we found that most of the MTurk participants reported watching the video stream of at least one of the sUAS that reported an event, while only about 20% reported viewing video streams from both sUAS. We observed a similar pattern in the study with domain experts because all (7/7) experts were able to validate the presence of a person in the video and two experts also detected the wrong autonomous action of the sUAS for

---
[2] https://tinyurl.com/UAV-Autonomy-Artifacts

Scenario V3. In case of V4, five of the seven domain experts managed to find the presence of the person in the video but were not sure about the details of other event. This also suggests that domain experts were more interested in "Person Detection" event than the "Mist" event.

A notable issue experienced primarily by MTurk participants was switching attention between the map view and the sUAS information panel. One participant stated that "*I was watching the map and trying to figure out which color was which drone compared from the right window to the left window*". A few MTurk participants from V4 also mentioned looking at the pictures in the explanation box (*Information Clustering*) instead of the raw video stream, observing that *"there were pictures showing the person that had been detected"*. Notably, in scenario V5, where we used visual communication to draw users' attention towards UAVs' request (see Fig 4d) for a safe landing, all domain experts except one showed awareness of the mechanical failure and said that they would have taken manual control of the sUAS as requested.

**Perceived Workload**
Monitoring multi-UAV missions requires human operators to analyze, comprehend and make decisions based on rapid information coming from multiple UAVs. Interpreting explanations can further increase the cognitive load of humans. Therefore, we utilized the Raw NASA-TLX (NASA-RTLX) to measure the perceived workload (Hart 2006) of MTurk participants. The NASA-RTLX measures perceptions of workload for physical, mental, and temporal demands as well as performance, effort, and frustration. As our task did not involve any physical activity, we did not ask our participants to report their perception of the physical load.

Figure 5 shows the categorical NASA-RTLX scores across all scenarios. Unsurprisingly, irrespective of the design of explanations, the mental demand, temporal demand, and effort were found to be higher when multiple events occurred at the same time and where at least one of the sUAS failed to respond correctly (V3 and V4). Ruff et al., (Ruff et al. 2004) found that the perceived workload of operators increases under autonomous operation compared to manual. Our results indicate that the workload further increases as the number of autonomous actions and associated explanations increase. The median workload for scenarios in V1 (66.0) and V2 (72.50) was comparatively lower than when multiple events occurred in V3 (86.0) and V4 (84.5) as can be observed from Figure 6. On the other hand, when the monitoring task did not require the participants to shift their focus from the map view to the video stream view in the right-hand side panel to perceive the environment, the median workload for simultaneous events in V5 (74.0) was found to be lower as compared to that of V3 and V4. This also suggests that the design of explanations must minimize the need to switch attention in order for users to understand autonomous actions.

In conclusion, both MTurk participants and domain experts demonstrated similar behavior regarding the perception of sUAS autonomy, automation bias, and partnership with sUAS. Further, In response to **RQ3**, we found that De-

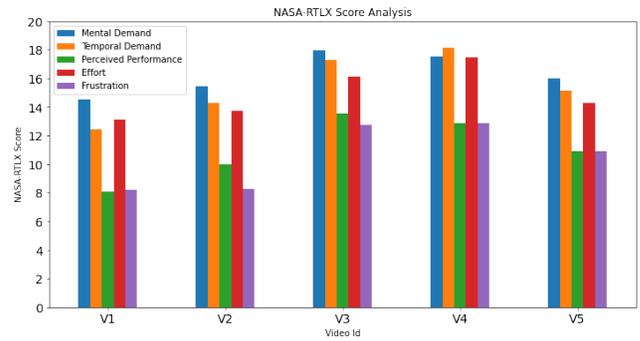
Figure 5: NASA-RTLX Score distribution of MTurk participants

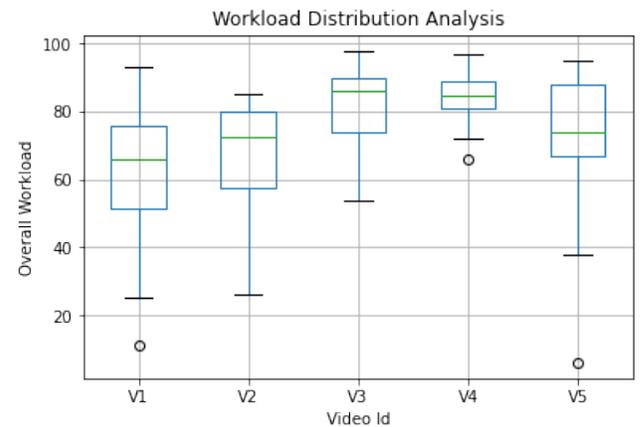
Figure 6: Overall Workload distribution of MTurk participants across all scenarios

sign Concepts such as Event-Action pairing, graphical animations, and information clustering improved SA whilst keeping the information and cognitive load low. Event-Action Pairing helped in conveying the explanation(s) in a concise format, reduced information overload, and helped our participants maintain a clear separation between the distinct autonomous actions of multiple sUAS resulting in better awareness of the situation. Graphical animations were useful for immediately drawing a user's attention in emergency cases such as mechanical failure in a sUAS. Information clustering allowed participants to evaluate the sUAS autonomy from the images in the explanation boxes, whilst minimizing the need to switch attention between views.

## Multi-sUAS UI Design Implications
### Adaptive Levels of Details
Our sUAS autonomy awareness analysis suggests that detailed explanations should be omitted when multiple sUAS encounter simultaneous events, meaning that the details to include in the explanations need to be adapted to support easy comprehension. In the context of process monitoring, Matkovic et al., (Matkovic et al. 2002) adapted the level of detail for effectively visualizing the reading from multi-instruments. Similarly, the level of detail in the sUAS explanations could be adapted based on multiple factors such as confidence of sUAS in performing the action autonomously, information already present in the UI, and the potential impact on the mission if the autonomous actions are unwarranted. For example, if one sUAS adapts its flight pattern in response to adverse environmental conditions, while another transitions to tracking mode in response to sighting a potential victim, it would make sense to provide more details in the explanation of the critical tracking decision than others in order to reduce the risk of information overload.

### Information Count

In our automation-bias analysis, we found that on average participants were able to correctly report four different details about the autonomy of sUAS and the surrounding environment, regardless of the number of events or the correctness of sUAS autonomous actions. Participants also paid attention to sUAS colors, and the direction in which some or all sUAS were moving. While the psychology literature reports that short-term memory of a human can hold between five and nine pieces of information (Miller 1994), we observed that this number was limited to four pieces of information associated with autonomy explanations, most likely due to the additional stress of a rapidly evolving environment, and the need to recall additional information related to recent events or the general context of the mission. This suggests that the UI should limit autonomy explanations to four critical pieces of information.

### Prioritize Explanations

When simultaneous events occur in the environment, the explanations should be prioritized according to the degree of uncertainty in the sUAS action and the consequence of those actions. The explanations of autonomous sUAS actions that could jeopardize the mission should receive the highest priority. Prioritizing the explanations may cause remote operators to remain unaware of some sUAS autonomous actions. However, such trade-offs seem necessary for the overall purpose of the mission. Therefore, the system designers need to carefully investigate the possible occurrence of events in various mission contexts to prioritize explanations.

### Event Validation

Finally, we observed that the participant's ability to maintain sufficient SA depends on their ability to discern whether the autonomous action of sUAS was justified in response to the current event, or even whether the event reported by the sUAS actually occurred. In our experiments, the problem of automation bias was repeatedly observed as users failed to notice errors made by sUAS. Therefore, when an sUAS has seemingly low confidence in its decision but its autonomy model supports execution of the action, the explanations should focus on providing information to help users to discern what actually happened and whether the autonomous action was justified. For instance, if an sUAS starts tracking a person with seemingly low confidence in the image detection, the explanation of such an autonomous action should encourage the user to inspect the video and confirm or refute the decision. We hypothesize that empowering the user to evaluate decisions leads to more correct outcomes than simply showing confidence scores. This strategy would increase the human cognitive load to interpret the correctness of the autonomous decision at the expense of protecting them from confirmation bias and a possible hazardous outcome. The UI designers should carefully consider such trade-offs when designing a multi-sUAS interface.

## Limitations and Future Work

We utilized simulated sUAS in our user studies to collect feedback from participants. While a controlled experiment in a real-world setting with several physical sUAS will be important for learning more about UI design trade-offs; our design setup allowed us to control the scenarios precisely, to engage a larger set of participants, and to explore many different design decisions. However, our study had certain limitations. Study participants were not able to interact directly with the prototype because we used prerecorded simulated videos. We plan to conduct an interactive study with additional expert users in the future to further validate our proposed design guidelines. Second, we only studied the impact of two simultaneous events and considered only four primary autonomous features. Further work is needed to design solutions that would be resilient to even more simultaneous events; however, the approaches we have identified through our study can be applied and evaluated in future real-world studies with physical sUAS.

## Conclusion

Since explanations from multiple sUAS are sometimes necessary, we emphasize the problem of information overload and its impact on human operators' situational awareness. We followed an iterative approach to design, evaluated an explainable UI for autonomous multi-sUAS human-on-the-loop systems, and collected feedback from inexperienced crowd workers and experienced sUAS operators. In particular, we focused on exploring the design trade-offs implicit to handling simultaneous autonomy explanations in a multi-sUAS environment. We explored several design choices for explaining the autonomous behavior of multiple sUAS and documented their impact on remote operators' situational awareness. Several design choices such as the Event-Action pairs, information clustering, and use of graphical animations over natural language in specific mission context were found to be effective for explaining autonomous actions. We have provided actionable guidelines for effectively explaining the autonomous behavior of multiple sUAS. We anticipate the findings may apply to other fleet monitoring applications, such as monitoring multiple robot taxis for safe operations.

## Acknowledgements

The work described in this paper is funded by the USA National Science Foundation grant CNS-1931962.